# Free space optical communication receiver based on a spatial demultiplexer and a photonic integrated coherent combining circuit

VINCENT BILLAULT,[1,*] JEROME BOURDERIONNET,[1] JEAN PAUL MAZELLIER,[1] LUC LEVIANDIER,[1] PATRICK FENEYROU,[1] ANAELLE MAHO,[2] MICHEL SOTOM,[2] XAVIER NORMANDIN,[3] HERVE LONJARET,[3] ARNAUD BRIGNON,[1]

[1]*Thales Research and Technology, 1 Avenue Augustin Fresnel, 91767 Palaiseau, France.*
[2]*Thales Alenia Space, 26 Avenue J.F. Champollion – B.P. 33787, 31037 Toulouse Cedex 1, France*
[3]*Thales LAS, 2 Avenue Gay Lussac 78995 Elancourt, France*
*\*vincent.billault@thalesgroup.com*

**Abstract:** Atmospheric turbulence can generate scintillation or beam wandering phenomena that impairs free space optical (FSO) communication. In this paper, we propose and demonstrate a proof-of-concept FSO communication receiver based on a spatial demultiplexer and a photonic integrated circuit coherent combiner. The receiver collects the light from several Hermite Gauss spatial modes and coherently combine on chip the energy from the different modes into a single output. The FSO receiver is characterized with a wavefront emulator bench that generates arbitrary phase and intensity patterns. The multimode receiver presents a strong resilience to wavefront distortions, compared to a monomode FSO receiver. The system is then used to detect an analog modulation of an optical beam through a random wavefront profile to mimic the transmission of a signal on a degraded optical link.

## 1. Introduction

Optical feeder links are currently considered as a promising technology to cope with the continuous growth in high data rate communication demand [1-2]. However, two mains factors limit the performances of optical communications with satellites: the low signal power delivered to the receiver and the atmospheric turbulence [3]. The low signal power and the weak signal-to-noise ratio at the receiver side practically limits the performances of the communication link. Besides, atmospheric turbulence distort the optical wavefront degrading the optical coherence of the laser beam [4] and causing signal fading. These combined effects reduce drastically the single mode fiber (SMF) coupling efficiency of the free space optical (FSO) receiver.

To preserve a good coupling efficiency in a SMF FSO receiver, the aperture size of the collection lens is limited by the correlation length of the atmospheric turbulence to tens of centimeters [5]. A first route to overcome this limitation, lies in adaptive optics. It mitigates the atmospheric distortion with deformable mirrors. However, the complexity of the control algorithm limits in practice the correction loop bandwidth [6]. Recently, different realizations of multichannel FSO receivers have been proposed with fast electronic signal recovery [7-8]. For multi-aperture (MA) receivers, the input wavefront is sampled by an aperture array to sum the energy from different positions of the wavefront. MA receivers are routinely used for radiofrequency systems [9], and MA FSO receivers have been presented to improve the optical feeder links efficiency [7]. However these architectures require a coherent receiver for each aperture and a synchronization of the channels in the electrical domain which impose strict constraints on the digital processing performances and limits practically the scalability to a high number of aperture. Alternatively, an original architecture of multichannel FSO receiver with a multimode fiber has been proposed to increase the collection area [8]: the light is coupled to a multimode fiber and a spatial demultiplexer projects the coupled light onto the multimode fiber's modes basis. Each projection is then converted into a Gaussian profile and coupled to a separate SMF. Similarly to the MA FSO receiver, the collection area is wider than a SMF but it requires a coherent receiver for each mode which limits the scalability of the system.

Recently, binary tree architectures (BTA) have been proposed to combine the light from an array of optical input beams to one output beam [10]. For FSO receivers, BTA avoid detecting the light out of each channel and subsequently adding the data signal in the electrical domain after re-synchronization process. The use of BTA with FSO receiver has been proposed in [11-12] for a few number of inputs with discrete elements. Practically, these systems show an enhancement of the signal to noise ratio and an also facilitate the demodulation step avoiding the active resynchronization of the channels. However, the setups with off-the-shelf elements are not viable for the operation with a large number of inputs, as the system size and total propagation losses increase with the number of modes [11].

Therein, photonic integrated circuits (PIC) exhibit attractive features for achieving the coherent combination of the multiple outputs of the receiver. PICs provide both optical and electronic functions with ultimate small footprint, and a remarkable thermal and mechanical stability [13-14]. The photonic integrated platform is also perfectly suited to scale to a high number of inputs with a wide flexibility to standard BTA [15] to fully reconfigurable photonic circuits [16], as the propagation on the chip through phase modulators and couplers can be as low as 0.5 dB/cm [17].

In this paper, we propose and demonstrate for the first time of our knowledge a complete FSO receiver for optical communications associating a spatial demultiplexer and a BTA on a PIC. This architecture thus combines the advantages of a BTA supporting multimode collection area and the high integration capability and robustness of PICs. In the first section, we describe the wavefront emulator that we developed in order to mimic phase and amplitude distortion and evaluate the receiver performances. The emulator is calibrated to selectively excite the demultiplexer eigenmodes in order to generate arbitrary phase and amplitude spatial beam profiles as a combination of the demultiplexer mode basis. In a second section we describe and characterize the combining PIC. In the last section the performances of the complete FSO receiver are evaluated.

## 2. Experimental setup.

The proposed receiver and wavefront emulator test bed are presented in Figure 1. The wavefront emulator (Fig. 1A) generates arbitrary amplitude and phase light field. The input profiles are decomposed on the Hermite Gauss (HG) basis by the spatial demultiplexer (Fig. 1B), and each projection is coupled to a SMF. Finally, the power of each fiber is coherently combined with a PIC on a single output beam (Fig. 1C).

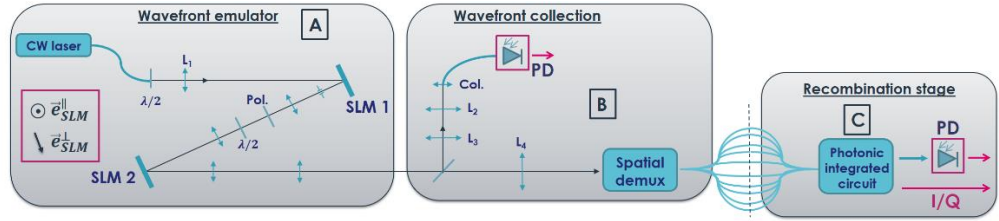

Fig.1. Experimental setup of the FSO receiver. Col : collimator. Pol : polarizer. PD : photodiode.
SLM : spatial light modulator. Focal lens : $L_1 = 30$ cm, $L_2 = 7.5$ cm, $L_3 = 30$ cm, $L_4 = 50$ cm.

The wavefront emulator is inspired from [18] and uses two nematic liquid crystal phase-only spatial light modulators (LCOS SLMs) for intensity and phase modulation of an input field. The light from a CW laser operating at 1.55 µm passes through a half-wave plate (HWP) and a collimating lens ($L_1$) to reach the $SLM_1$. A 4f afocal system (with unit magnification) images the plane of SLM$_1$ to that of SLM$_2$. An iris, a polarizer and a HWP are placed between the two SLMs. The two SLMs modulate spatially the phase of the light for only one polarization direction $\vec{e}_{SLM}^{\parallel}$ orthogonal to the plane of Figure 1. The first HWP controls the polarization of the light that addresses SLM$_1$, in order to have the light linearly polarized along $\vec{e}$, oriented at 45° from $\vec{e}_{SLM}^{\parallel}$. With a polarizer in the same polarization direction $\vec{e}$, the SLM$_1$ associated to the polarizer acts like an intensity spatial modulator. The second HWP orients the light polarization direction to $\vec{e}_{SLM}^{\parallel}$ so that the SLM$_2$ acts like a phase only spatial modulator. By applying respectively phase patterns $\varphi_1(x,y)$ and $\varphi_2(x,y)$ to SLM$_1$ and SLM$_2$, the electric field $E(x,y)$ at the wavefront emulator output writes [18]:

$$E(x,y) = A_0 \cos\left(\frac{\varphi_1(x,y)}{2}\right) \exp\left(i\frac{\varphi_1(x,y) + 2\varphi_2(x,y)}{2}\right), \quad (1)$$

with $A_0$ the amplitude of the input wavefront on the SLM$_1$. So, by changing the phase patterns written onto SLM$_1$ and SLM$_2$, we manipulate the amplitude and phase of the wavefront independently at the same time. Figure 2 shows several profiles of HG modes generated with the wavefront emulator imaged on a camera (not shown in Fig. 1). The camera is imaging the plane of SLM$_2$ through a 4f system (with a magnification of 0.7).

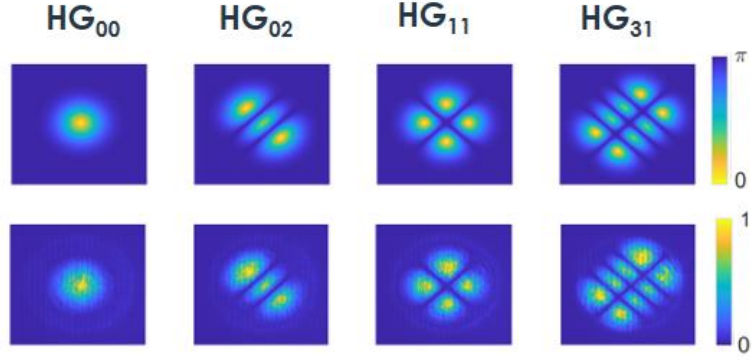

Fig. 2. Example of generated HG modes. Patterns display on SLM$_1$ (up) and image with the camera (bottom).

The multichannel FSO receiver (CAILabs: Tilba) is a mutli-plane light conversion (MPLC) module that takes a FSO wavefront for input and returns its projection on a 15 HG modes basis, on 15 separated SMF. The modes basis at the output is given by the MPLC (in our case the HG modes constitute a complete basis set of orthonormal functions that are therefore suited to the wavefront emulator). We first characterize the transmission of the MPLC module by generating consecutively the different HG modes with the wavefront emulator. For a given mode HG$_{x,y}$, we measure the free space input power ($P_{x,y}^{in}$) and the ouput power for each fiber (the 15 $P_{x',y'}^{out}$) with a photodiode array. Figure 3 shows the transmission matrix $P^{in}/P^{out}$ of the MPLC, and the Table 1 sums up the MPLC characteristics. For each mode, we calculate the insertion losses with $P_{x,y}^{out}/P_{x,y}^{in}$ and the cross talk with $P_{x',y' \neq x,y}^{out}/P_{x,y}^{in}$ of the MPLC in a receiver configuration. The transmission matrix is quasi-diagonal and the mean cross talk is very low ($< -19$ dB), making this architecture suitable for the reception and decomposition of distorted wavefronts. The mean insertion loss (for the 15 HG modes) of the MPLC when measured in the other direction, i.e. from the single mode input to the multimode output, is 1 dB. The higher insertion loss per mode we obtain ($\sim 3.6$ dB) comes mainly from the mismatch between the HG modes generated via the wavefront emulator and the actual eigenmodes of the MPLC.

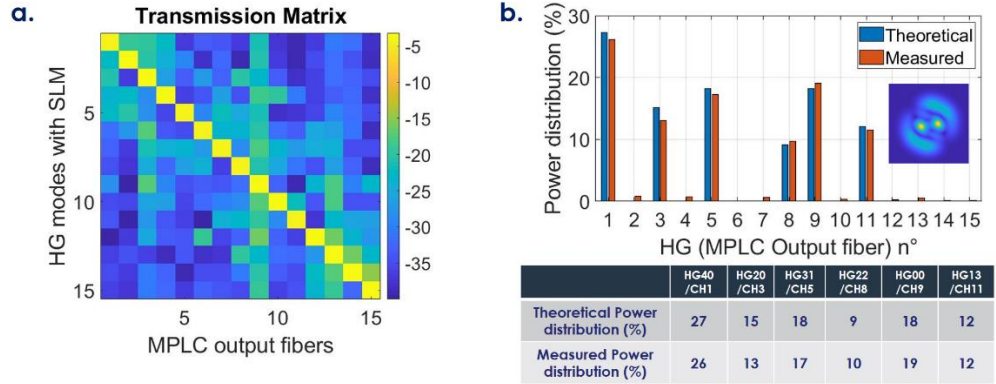

Fig. 3. (a) Transmission matrix of the MPLC (logarithmic scale) (b) Output power distribution (%); theoretical (blue) and measured via the photodiode array (red), inset : corresponding pattern on SLM$_1$.

Table 1. MPLC characteristics for HG$_{x,y}$: insertion losses (IL) and mean cross talk (MCT) in dB

| HG$_{x,y}$ | 40 | 30 | 20 | 10 | 31 | 21 | 11 | 22 |
|---|---|---|---|---|---|---|---|---|
| IL | -3.8 | -3.5 | -3.2 | -3.6 | -4.0 | -3.9 | -3.8 | -4.2 |
| MCT | -22 | -23 | -24 | -22 | -22 | -21 | -21 | -23 |

| HG$_{x,y}$ | 00 | 12 | 13 | 01 | 02 | 03 | 04 |
|---|---|---|---|---|---|---|---|
| IL | -3.5 | -3.4 | -4.0 | -3.7 | -3.2 | -3.2 | -3.6 |
| MCT | -21 | -21 | -22 | -19 | -21 | -20 | -23 |

To verify the MPLC module linearity, we generate a wavefront corresponding to a combination of 6 HG modes with random phase and power distribution. As the insertion losses slightly vary among the HG modes, we normalize each mode by the corresponding eigenvalue of the MPLC transmission matrix. The table in Fig.3b shows the theoretical random power distribution of the 6 modes. Figure 3 presents the output power distribution of the 6 fibers measured with the photodiode array. The output power distribution is in excellent agreement with the theoretical distribution. The small differences that emerge come from the phase flicker of the SLMs and the residual cross-talk of the MPLC.

### 3. Coherent combining photonic integrated circuit

We designed a photonic integrated BTA to combine the energy from the 15 output fibers of the MPLC. It has been fabricated within IMEC's (Belgium) Silicon On Insulator (SOI) photonics foundry through multiprojects wafers shuttle and packaged by PHIX (Netherlands). The operating principle of the BTA is described in Figure 4. The PIC is divided in 14 unit cells, each cell being composed of two inputs (with a phase shifter in one of them), a Mach-Zehnder interferometer between the two inputs (with a phase shifter in one arm) and two outputs [11-19]. A photodiode at one of the cell outputs detects the interference intensity. By sweeping the voltage applied to the cell thermo-optical phase shifters, one can minimize the current on the photodiode. When it is minimum, the power at the complementary output of the interferometer is maximum, i.e. both amplitude and phase imbalance have been corrected and the power from the two inputs of the cell are equalized and coherently combined. A Nelder Mead optimization algorithm [20] has been implemented to minimize cell by cell the current of the photodiodes by adjusting the phase shifters driving voltages starting from the cells at the input side of the PIC, to the final output cell [21]. With the associated electronics (transimpedance amplifiers and amplifiers with tunable gain), the sensitivity of the detection is lower than 1 nW which is perfectly suited to follow long-term variably low coupling efficiency. The combined beam can be routed for direct detection (either on an on-chip photodiode or coupled outside the chip) or for phase and quadrature (I/Q) coherent detection with an on-chip 90° coupler and balanced detectors [22].

The optical coupling loss from fiber to waveguide is estimated to ∼ 4dB / interface using shunt waveguides in the PIC (not represented). We measured a total insertion loss of 7dB at 1535 nm from any of the 15 fiber inputs to the on-chip direct or (I/Q) coherent detection. The 7dB breaks out in 4dB coupling loss plus 5 × 0.6dB for each of the 5 cascaded Mach-Zehnder interferometers in the total optical path of the PIC (see Fig. 4a). When the light is routed outside the chip, the insertion loss is increased by ∼ 4dB due to the waveguide to fiber coupling.

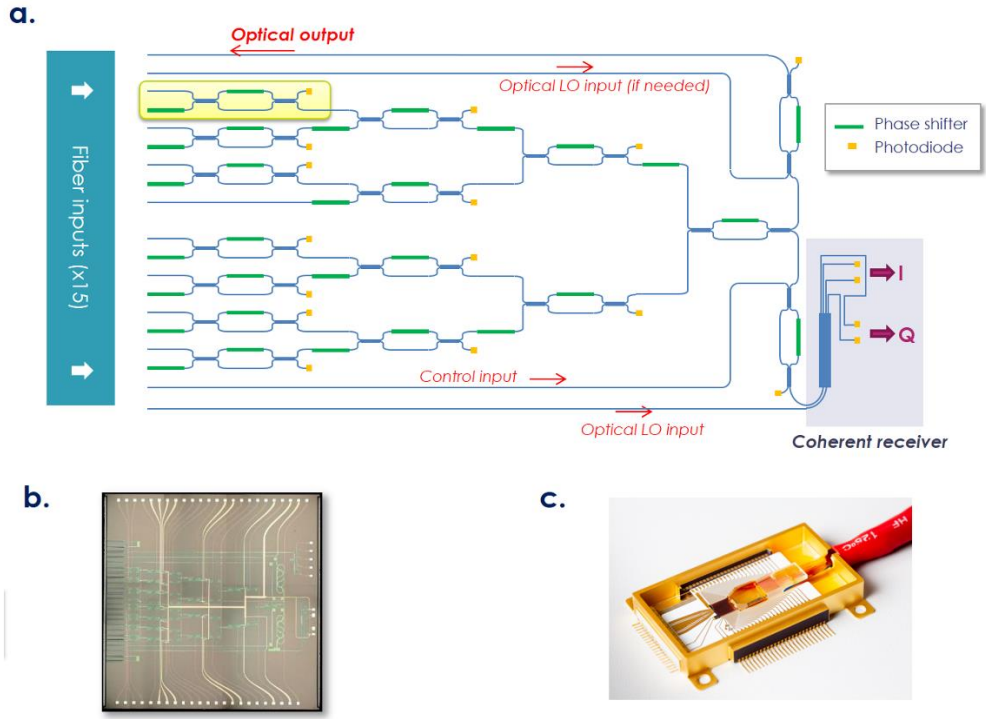

Fig. 4 (a) Scheme of the PIC. A unit cell is represented in yellow (b), (c) pictures of the PIC and packaged device.

To validate the operating principle of the PIC combiner, the optimization algorithm is implemented for optical waves from 6 fiber inputs as a first proof-of-concept experiment. Figure 5 shows the evolution of the combined power without and with the feedback loop. The bundle of fibers coupled to the PIC is squeezed manually to generate different phase perturbations in the fibers. In open loop configuration, the phase variations between the different inputs cause large variations of the optical power. As soon as the feedback loop is switched on, the combined power reaches and stays to a maximum corresponding to the total coherent combination of the inputs. The observed residual fast fluctuations are due to the 2 pi phase jumps of the phase shifters (50 µs time scale). The perturbations induced to the system were not quantified nor related to any real FSO conditions, and were intended here to assess he system ability to maintain the coherent combination under dynamic phase fluctuations conditions. The bandwidth of the loop correction is intrinsically limited by the dynamic of the thermooptic phase shifters that we used (typically ∼ 1-10 kHz). However, this bandwidth is theoretically sufficient to mitigate atmospheric turbulence, as the perturbation bandwidth is typically below 1 kHz [23-24].

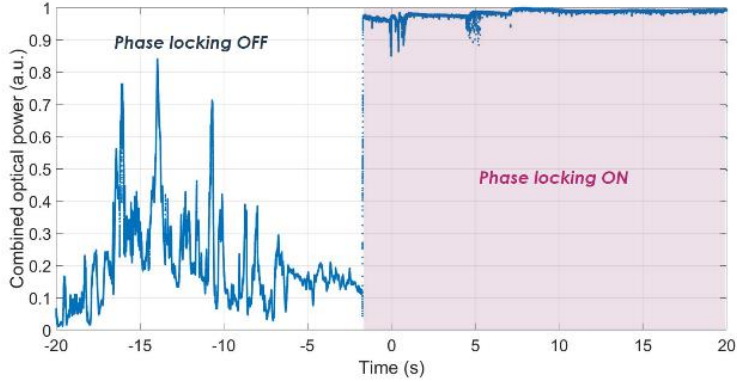

Fig. 5 Combined optical power at the PIC output with/without phase locking.

## 4. Experimental results.

To characterize the efficiency of the complete FSO emulator and receiver (MPLC and BTA on the PIC), we generate arbitrary wavefronts as combination of HG modes with the setup of Fig. 1, and we detect and combine the power from the different modes. To compare our system to conventional FSO receivers, we add a second collection arm (cf. Fig. 1B) composed of a collimator and a SMF. An afocal system (with a magnification of 0.25) adapts the size of the wavefront without any spatial shaping to the collimator to optimize the coupling efficiency. We apply different waveform profiles with the wavefront emulator and compare the coupling efficiencies for the two receivers. We changed the CW laser wavelength to 1535 nm to optimize the efficiency of the coherent combiner: the coupling loss to the PIC is wavelength independent ($\sim -4$dB) in the C band, but the propagation loss of the light on the PIC depends significantly of the wavelength (mainly because of the multimode interference couplers).

For the monomode receiver, the coupling efficiency $\eta_{\text{mono}}$ corresponds to the ratio of the coupled power in the output single mode fiber $P_{\text{out}}$ to the free space input power $P_{\text{in}}$ (cf. Fig 6.a). For the multimode receiver, we defined the efficiency $\eta_{\text{PIC}}$ as the ratio of the combined power on-chip $P_{\text{out}}$ to the sum of the power in each fiber after the MPLC $P_{\text{in1}}$ (cf. Fig 6. b). Power-on-chip is considered here, illustrating the additional benefit of the PIC which incorporates the detectors, thus avoiding out-coupling losses. The set of beam profiles tested corresponds to a combination of 6 HG modes with random amplitudes and phase (The six HG modes used are the modes $HG_{22}$, $HG_{31}$, $HG_{00}$, $HG_{20}$, $HG_{13}$, $HG_{40}$).

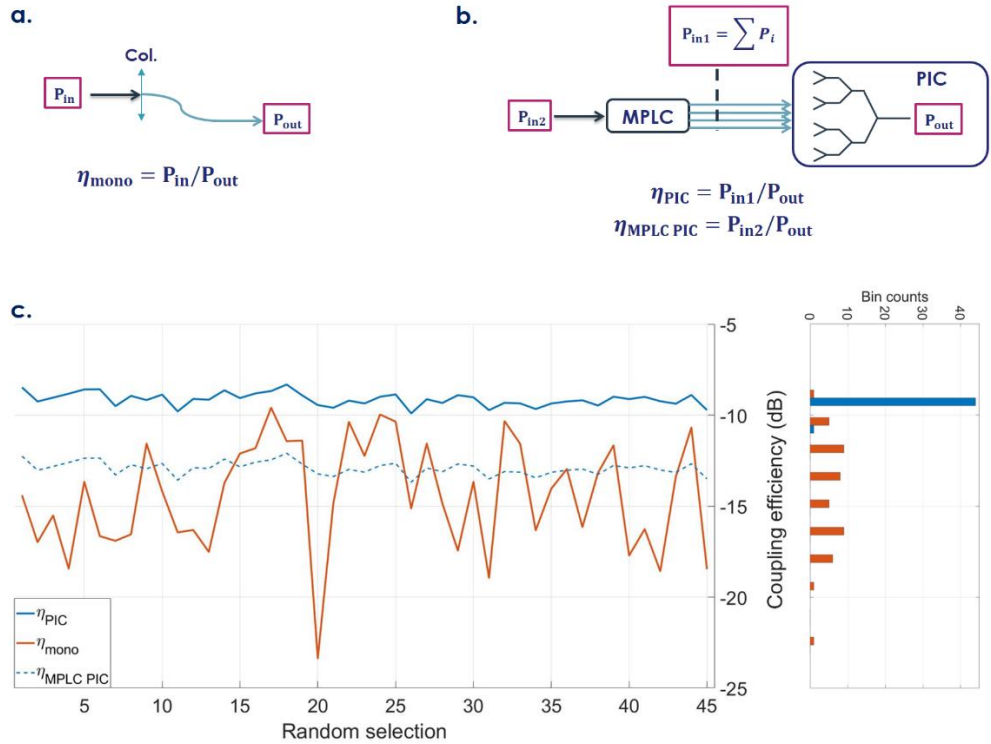

Fig. 6 (a) Scheme of the efficiency measurement (a) for the monomode FSO receiver (b) for the MPLC + PIC FSO receiver. (c) Left : Coupling efficiency for the monomode FSO receiver and the MPLC + PIC FSO receiver for random beam. Right : histogram of the coupling efficiency

For no spatial shaping on the SLM, the coupling efficiency for the SMF FSO receiver is equal to 4.6dB. The coupling efficiency for the SMF FSO receiver is higher than for the MPLC and PIC FSO receiver, mainly due to the propagation and coupling loss of our present PIC device (~ 7 dB). However, for random patterns, the coupling of the SMF FSO receiver strongly varies as the coupling of the spatial demultiplexer and PIC FSO receiver remains almost constant (see Figure 6.c). This shows that the latter is much more resilient than classical FSO receivers to phase and amplitude perturbations. This property is extremely valuable for telecommunications, as atmospheric perturbations rapidly change the phase and energy distributions between the modes. Besides, the efficiency of the PIC coherent combiner is higher than the FSO receiver for every random pattern tested. Even when considering also the loss of the MPLC (which are overestimated due to the wavefront emulator, as seen in section 2), the efficiency $\eta_{\text{MPLC\_FSO}}$ (corresponding to the ratio of the combined power on-chip $P_{\text{out}}$ to the free space input power $P_{\text{in}}$, see Fig. 6.b, dashed blue line), is higher in average than the efficiency of the monomode FSO receiver.

Finally, to detect radiofrequency modulations applied to the electric field with the full setup (wavefront emulator and multimode receiver), we used the on-chip coherent receiver. The coherent receiver is composed of a 2x4 multimode interference (MMI) 90° hybrid coupler and two pairs of balanced photodiodes (cf. Fig. 4. a) to generate the phase and quadrature (I/Q) component of the radiofrequency field. We characterize the power and phase imbalance between the 4 outputs of the MMI coupler with an unbalanced Mach-Zehnder interferometer test structure (MZI, cf. scheme of Fig. 7.a) [25], positioned at a corner of the PIC. By sweeping the wavelength of the MZI input, we determine the relative phase offsets and power imbalance of the 2x4 MMI hybrid coupler at the output ports (cf. Fig. 7.a). The MMI outputs are shifted from respectively 0, 82, 184 and 268° with a 0.4 dB power imbalance between the outputs.

As a preliminary evaluation of the on-chip I/Q receiver, we used an acousto-optic frequency shifter placed upstream of the wavefront emulator input to generate an analog frequency modulation of the optical link at 80 MHz. To mimic a degraded microwave photonic link, a random HG modes combination extracted from Fig. 6 is applied to the SLMs. Therefore, the MPLC output corresponds to an excitation of 6 fibers, carrying each the 80 MHz analog frequency modulation. The relative true time delays mismatch between each channel ($\delta t$) must been compensated in order to use modulated signal with the BTA ($\delta t \ll 2/BW$, with BW the modulation bandwidth). With BW = 80 MHz, the true time delay must be much lower than 25 ns (7.5 m of free space propagation). For that modulation bandwidth, the multimode FSO receiver presented in Fig. 1 is precise enough. The on-chip binary tree architecture with the Nelder Mead algorithm combines the energy of the 6 fibers in one single beam that is routed to the on-chip coherent receiver. The analog modulation is detected with respect to the CW laser that is coupled to the chip. Figure 7.b shows the temporal profile of the I/Q detection at the coherent receiver output. The analog phase modulation (at 80 MHz) is successfully reconstructed after free space propagation through a random perturbation, collection with the MPLC and coherent recombination. This result validates the use of the spatial demultiplexer and PIC based FSO receiver for the balanced detection of modulated waveform.

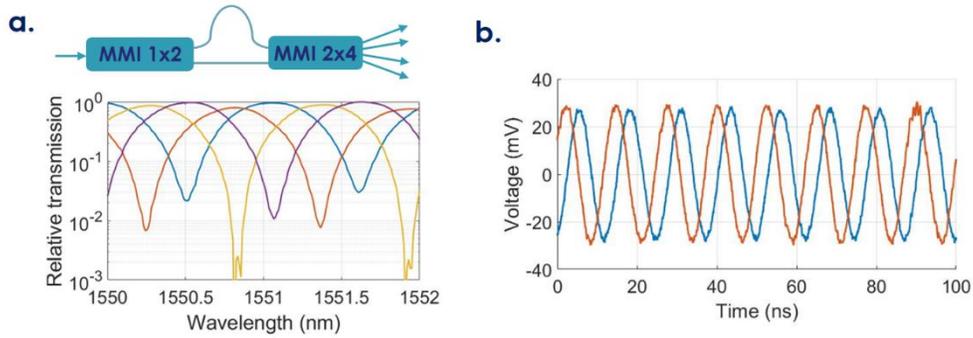

Fig. 7 (a) Characterization of the 2v4 MMI coupler: scheme of the MZI (up) and relative transmission of the 4 MMI outputs versus the input wavelength of the MZI (bottom) (b) Outputs of the on-chip coherent receiver (red : I, blue Q) with a 80 MHz modulation applied to the wavefront emulator input.

## 5. Conclusion.

We have proposed and demonstrated a proof-of-concept FSO receiver based on a spatial demultiplexer and a PIC coherent combiner. We have developed a wavefront emulator to test the receiver with arbitrary amplitude and phase patterns. The FSO receiver has proven the efficient coherent combination of 6 fibers for different random inputs. Compared to a conventional SMF FSO receiver, we have showed that our system is much less sensitive to phase and amplitude perturbations. In a final experiment, we validate the FSO receiver for the on-chip detection of a modulated input wavefront. This first characterization of the setup shows the great potential of this complete architecture, from the emulator bench to the BTA on the PIC. This setup will be further improved with a fine tuning of the delays between the different channels (with tunable optical delay line) to use the FSO receiver with modulated signals with an higher modulation bandwidth and more complex frequency modulation scheme (OOK, QPSK, DQPSK, …). The implementation of realistic atmospheric perturbation scenario (with the projection of simulated distorted wavefront on the HG basis) is also currently considered.

As a conclusion, the implementation of FSO receivers based on a spatial demultiplexer with coherent combination presents several advantages compared to standard SMF FSO receiver: The architecture shows a strong resilience to random phase and intensity perturbation, as the coupling efficiency does not depend on the energy distributions among the spatial modes. As a consequence, this systems ought to be less sensitive to atmospheric turbulence. Then, the system is largely scalable, as MPLCs can project and convert wavefronts into a large number of spatial modes, and integrated photonic platforms allow to encompass a large number of photonic devices on the same chip. Moreover, since insertion loss in the PIC are mostly dominated by coupling loss, increasing the channel count will have a limited impact on the total insertion loss. For instance, since we estimated in section 3 the loss of one combining cell to ~0.6dB, then doubling the number of input channels would only add 0.6dB to the total insertion loss. The scalability of the architecture to a large number of modes can provide significant benefit for FSO communication receivers: as the spatial demultiplexer collects a large number of spatial modes and have a high numerical aperture, the collection area is larger than standard SMF FSO receiver. The aperture diameter of the collection lens with this system could therefore become higher than the atmospheric turbulence correlation length and so increase the received signal power. The present approach also opens up prospects for upcoming hybrid PIC platforms combining SOI, SIN and InP, offering ultra-low loss waveguides, high bandwidth phase shifters and high performance photodiodes, which would benefit the number of recombined channels.

**Acknowledgment**

We thank Arnaud Le Kernec for useful discussions. This work was partially supported by the VERTIGO project from the European Union's Horizon 2020 research and innovation program under grant agreement No. 822030.

**Disclosure**

The authors declare no conflicts of interest.

**Data Availability**

Data underlying the results presented in this paper may be obtained from the authors upon reasonable request.

**References**

1. R. Saathof, R. den Breeje, W. Klop, S. Kuiper, N. Doelman, F. Pettazzi, A. Vosteen, N. Truyens, W. Crowcombe, J. Human, I. Ferrario, R. M. Calvo, J. Poliak, R. Barrios, D. Giggenbach, C. Fuchs, and S. Scalise, "Optical technologies for terabit/s-throughput feeder link," in Proceedings of IEEE International Conference on Space Optical Systems and Applications (IEEE, 2017), pp. 123-129,
2. A. Le Kernec, L Canuet, A. Maho, M. Sotom, D. Matter, and L. Francou, "Optical feeder links for high throughput satellites and the H2020 VERTIGO project," COAT-2019-workshop (Communications and Observations through Atmospheric Turbulence: characterization and mitigation) (2019).
3. A. K. Majumdar and J. C. Ricklin, *Free-space laser communications: principles and advances,* Vol. 2. (Springer-Verlag, 2008).
4. Y. Dikmelik and F. M. Davidson, "Fiber-coupling efficiency for free-space optical communication through atmospheric turbulence," Appl. Opt. **44**(23), 4946–4952 (2005).
5. S. Shaklan and F. Roddier, "Coupling starlight into single-mode fiber optics," Appl. Opt. **27**(11), 2334-2338 (1988).
6. L. Rinaldi, V. Michau, N. Védrenne, C. Petit, L. M. Mugnier, C. Lim, J. Montri, L. Paillier, and M. Boutillier, "Sensorless adaptive optics for optical communications," Proc. SPIE **11678**, (2021).
7. D. J. Geisler, T. M. Yarnall, M. L. Stevens, C. M. Schieler, B. S. Robinson, and S. A. Hamilton, "Multi-aperture digital coherent combining for free-space optical communication receivers," Opt. Express **24**(12), 12661-12671 (2016).
8. M. Arikawa and T. Ito, "Performance of mode diversity reception of a polarization-division-multiplexed signal for free-space optical communication under atmospheric turbulence," Opt. Express **26**(22), 28263-28276 (2018).
9. V. Jamnejad, J. Huang, B. Levitt, T. Pham, and R. Cesarone, "Array antennas for JPLNASA deep space network," in Proceedings of IEEE Aerospace Conference Proceedings 2, 911–921, (2002).
10. D. A. B. Miller, "Self-aligning universal beam coupler," Opt. Express **21**(5), 6360-6370 (2013).

11. Y. Yang, C. Geng, F. Li, G. Huang, and X. Li, "Multi-aperture all-fiber active coherent beam combining for free-space optical communication receivers," Opt. Express, **25**(22), 27519-27532, (2017).
12. B. Zhang, J. Sun, C. Lao, C. H. Betters, A. Argyros, Y. Zhou, and S. G. Leon-Saval, "All-fiber photonic lantern multimode optical receiver with coherent adaptive optics beam combining," arXiv preprint arXiv:2105.09516, (2021).
13. C. Scarcella, J. S. Lee, C. Eason, M. Antier, J. Bourderionnet, C. Larat, E. Lallier, A. Brignon, T. Spuesens, P. Verheyen, P. Absil, R. Baets, and P. A. O'Brien, "Plat4m Progressing silicon photonics in Europe," Photonics. **3**(1), 1-10 (2016).
14. C. G. H. Roeloffzen, L. Zhuang, C. Taddei, A. Leinse, R. G. Heideman, P. W. L. van Dijk, R. M. Oldenbeuving, D. A. I. Marpaung, M. Burla, and K. J. Boller, "Silicon nitride microwave photonic circuits," Opt. express **21**(19), 22937-22961 (2013).
15. D. A. B. Miller, "Analyzing and generating multimode optical fields using self-configuring networks," Optica, **7**(7), 794-801 (2020).
16. W. Bogaerts, D. Pérez, J. Capmany, D. A. B. Miller, J. Poon, D. Englund, F. Morichetti, and A. Melloni, "Programmable photonic circuits," Nature, **586**(7826), 207-216, 2020.
17. S. K. Selvaraja, P. De Heyn, G. Winroth, P. Ong, G. Lepage, C. Cailler, A. Rigny, K. Bourdelle, W. Bogaerts, D. Van Thourhout, J. Van Campenhout, and P. D. VanThourhout, J. Van Campenhout, and P. Absil, "Highly uniform and low-loss passive silicon photonics devices using a 300mm CMOS platform," in Optical Fiber Communication Conference 2014, United States of America, paper Th2A.33 (2014).
18. L. Zhu and J. Wang, "Arbitrary manipulation of spatial amplitude and phase using phase-only spatial light modulators". Sci. Rep. **4**(1), 1-7(2014).
19. A. Annoni, E. Guglielmi, M. Carminati, G. Ferrari, M.Sampietro, D. A. B. Miller, A. Melloni, and F. Morichetti. "Unscrambling light—automatically undoing strong mixing between modes," Light Sci. Appl. **6**(12), 2017.
20. J. A. Nelder and R. Mead, "A simplex method for function minimization," Comput J. **7**(4), 308-313 (1965).
21. K. Choutagunta, I. Roberts, D. A. B. Miller and J. M. Kahn, "Adapting Mach--Zehnder mesh equalizers in direct-detection mode-division-multiplexed links," J. Light. Technol., **38**(4), 723-735 (2019).
22. J. Zhang, J. Verbist, B. Moeneclaey, J. Van Weerdenburg, R. Van Uden, H. Chen, J. Van Campenhout, C. Okonkwo, X. Yin, J. Bauwelinck, and G. Roelkens, "Compact low-power-consumption 28-Gbaud QPSK/16-QAM integrated silicon photonic/electronic coherent receiver," IEEE Photon. J. **8**(1), 1-10 (2016).
23. M. Toyoshima, S. Yamakawa, T. Yamawaki, K. Arai, M. Reyes, A. Alonso, Z. Sodnik, and B. Demelenne, "Ground-to-satellite optical link tests between Japanese laser communications terminal and European geostationary satellite ARTEMIS," Proc. SPIE **5338**, 1–15 (2004).
24. M. R. Garcia-Talavera, S. Chueca, A. Alonso, T. Viera, and Z. Sodnik, "Analysis of the preliminary optical links between ARTEMIS and the Optical Ground Station," Proc. SPIE **4821**, 33–43 (2002).
25. K. Voigt, L. Zimmermann, G. Winzer, H. Tian, B. Tillack, K. Petermann, "C-Band Optical 90° Hybrids in Silicon Nanowaveguide Technology," IEEE Photon. Technol. Lett. **23**(23), 1769-1771 (2011).